# A Critical Review on the Electromigration Effect, the Electroplastic Effect, and Perspectives on Effect of Electric Current upon Alloy Phase Stability

Yu-chen Liu[1] and Shih-kang Lin[1,2,3,*]

[1] Department of Materials Science and Engineering, National Cheng Kung University, Tainan city 70101, TAIWAN; [2] Center for Micro/Nano Science and Technology, National Cheng Kung University, Tainan city 70101, TAIWAN; [3] Hierarchical Green-Energy Materials (Hi-GEM) Research Center, National Cheng Kung University, Tainan 70101, TAIWAN

*E-mail: linsk@mail.ncku.edu.tw

## Abstract

The electronic interconnections in the state-of-the-art integrated circuit (IC) manufacturing are scaled down to the micron or sub-micron scale. This results in a dramatic increase in the current density passing through interconnections, so the electromigration (EM) effect plays a significant role in the reliability of products. Although thorough studies and reviews of EM effects have been continuously conducted in the past 60 years, some parts of EM theories lack clear elucidation of the electric current-induced *non-directional* effects, including the electric current-induced phase equilibrium changes. This review article is intended to provide a broad picture of electric current-induced lattice stability change and to summarize the existing literature on EM-related phenomena, EM-related theoretical models, and relevant effect of the electroplastic (EP) effect in order to lead to a better understanding of the electric current-induced effects on materials. This article also posits that EM is either part of the EP effect or shares the intrinsic electric current-induced plastic deformation associtated with the EP effect. This concept appears to contribute to the missing parts of the EM theories.

**Background**

Electric current is intimately related to human life. During the past hundred years, thorough theoretical models have been proposed for the electric current-induced effect, such as the thermoelectric effect, and the Joule heating effect. Currently, the technology node of state-of-the-art integrated circuit (IC) manufacturing can be scaled down to the nano scale. This leads to the interconnections being scaled down to the micron or sub-nano scale [1-3]. The scaling-down tremendously multi-functionalizes modern electronic devices, and allows continuous reductions in their feature size. However, the scaling-down has drawn attention to new effects induced by electric current due to increases in the current density passing through the interconnections, and resulted in significant reliability concerns [2, 4-10]. Therein, electromigration (EM) effect has caused significant reliability concerns in modern industrial electronic devices due to the formation of voids and hillocks [11], as well as the polarity effect [12, 13]. We call this EM-induced voids/hillocks formation and polarity effect *directional* effects because the atomic diffusion often follows the direction of the electron flow, which results in the formation of voids and hillocks as well as variations in the interface growth rate on opposite sites. The existing theories have successfully explained the EM-induced voids/hillocks formation and polarity effect [14].

Despite the success of EM theories, phenomena induced by electric current that have been found more recently cannot be thoroughly understood using the existing EM theoretical models [14-16]. For example, electric current is found to induce the alloy supersaturation effect [17, 18], the non-polarity effect [19], lattice deformation [20-26], electrorecrystallization [27], and grain rotation [28]. Unlike the voids/hillocks formation and the polarity effect, these phenomena are found non-directional, which means the effects do not follow the direction of electron flow and thus cannot be well explained by the existing EM models. We call them the *non-directional* effects. The missing parts of the existing EM theory motivated this article as a review of the existing literature on EM-related phenomena, EM-related theoretical models, and relevant electric current-induced effects such as the electroplastic (EP) effect. This study is an attempt to summarize all the EM- and EP-induced effects under a broad picture of electric current-induced lattice stability change. To be more specific, the existing literature either from EM or EP research points out that electric current will induce a lattice deformation, which is believed to be induced by atomic diffusion in EM and by the electron-dislocation interaction in EP. Based on the literature, an idea is proposed herein suggesting that either EM is part of the EP effect or shares the intrinsic electric current-induced plastic deformation induced by the EP effect. This idea may rationalize the electric current-induced phase equilibria change as well as the non-polarity effect.

First, the existing models for the EM driving force are reviewed, and the missing parts of these models are pointed out. The model of EM occurrence is then reviewed, where it is pointed out that the occurrence of EM is not only the function of the intrinsic resistivity property of materials but also that of the strip length, which is the so-called the Blech effect. The concept of EM-induced stress/deformation (e.g. back stress model) is introduced to elucidate the occurrence of EM. Early stages (i.e. before EM occurs) of lattice deformation found in the EM tests is reviewed, and correlations between the occurrence of EM and the mechanical properties of materials are pointed out. The relevant EP effect is then reviewed to reveal the intrinsic electric current-induced lattice deformation. Changes in electric current-induced phase equilibria and interfacial reactions are reviewed in the final section. Overall in this study, the missing parts of EM theories are revealed to clearly elucidate the *non-directional* effects.

**Theories of the electric current-induced force on ions (driving force for EM)**

Generally, the driving force of electromigration comprises two components, one of which is the direct force $F_d$ from the external electric field causing the columbic force, and the other of which is the electron wind force $F_w$. The general equation to formulate the driving force $F$ of EM can be written as Eq. (1):

$$F = F_d + F_w = (z_d + z_w)e\rho j = z^* e\rho j \qquad (1),$$

where $z_d, z_w, z^*$ are the effective charges in the direct force, electron wind force, and the net force, respectively. $e$ is the fundamental charge, $\rho$ is the resistivity, and $j$ is the current density. The effective charge sets the scale of the driving force and is one of the most important quantity by which to evaluate EM effect since it can be directly measured via experimental methods [14, 29]. The existing theoretical models tackled a fundamental understanding of $F_d$ and $F_w$. Thorough reviews on the existing theoretical models for EM have been provided elsewhere [15, 29-32]. In this section, we review four representative classic models and recently-proposed models, and highlight their potential issues.

*Semi-ballistic model*

Huntington *et al.* conducted a marker motion experiment in gold under an electric current, e.g. 30A and 60A, at high temperatures, e.g. 830-1020°C, and proposed the ballistic model for the EM effect [14]. They considered the momentum transfer per unit time to a point defect in a current-conducting metal. The electrons were considered to be scattered by the defect alone, which was decoupled from the lattice. The scattering process took place without the creation or annihilation of phonons. The x-component of the momentum transfer per unit time per unit volume to the defects $\frac{dM_x}{dt}$ is shown in Eq. (2):

$$\frac{dM_x}{dt} = -\left(\frac{1}{4\pi^3}\right)^2 \iint \frac{m_0}{\hbar}\left(\frac{\partial E}{\partial k'_x} - \frac{\partial E}{\partial k_x}\right) \times f(k)\{1 - f(k')\}\, W_d(k, k')\, dk'dk \qquad (2),$$

where $m_0$ is the mass of electron, $\frac{\partial E}{\hbar \partial k_x}$ is the group velocity of electron, $f(k)$ is the distribution function of electrons in k-space, and $W_d(k, k')$ is the transition probability function per unit time that the electron in state *k* will jump to state *k'* by virtue of its interaction with the defects. The momentum transfer between the electron and the defects is the mass times the group velocity. The problem with this model is that it is only applicable for an isolated defect in free electron gas. In real metals, the complicated band structure effect should be considered. The other concern is that not all the momentum lost by the electrons goes to the ion of interest [15]. The momentum lost may go to the lattice or neighboring scatters. Despite the fact the ballistic model is not general, it still formulates a pioneer physical picture of the driving force for EM.

*Charge polarization model*

Bosvieux and Friedel proposed a charge polarization model to approach the EM effect [16]. The charge polarization was more generalized and enabled a number of later modeling studies [15, 33-36]. Different from the momentum transfer proposed by the semi-ballistic model, the charge polarization model considered the electron wind force arising from the perturbed electron density in the vicinity of the bare defect complex, e.g. the combination of a bare ion with a vacancy in its vicinity. The perturbed electron density under electric field was assumed to come from two sources: (1) Arising from the incident electrons scattered by the defect complex, and (2) arising from the electrostatic electric field, including the polarization response. The general form of the electron wind force $F_w$ is therefore formulated as Eq. (3):

$$F_w = -\int \delta n(r) \nabla_R v_o(r - R) d^3r \qquad (3),$$

where $\delta n(r)$ is the perturbed electron density for any defects to generalize the formula to a realistic system, and $v_o$ is the bare potential for the defects. In order to obtain the $\delta n(r)$, one has to perform the self-consistent calculation to solve the Schrödinger equation. The total self-consistent potential is shown in Eq.(4):

$$v_{total}(r) = v_o(r) + v_{sc}(r) \qquad (4),$$

where $v_o(r)$ is the bare columbic potential of ion, and $v_{sc}(r)$ is the screening potential. The change in the electron density comes from the screening potential term, $v_{sc}(r)$. The self-consistent total potential and wavefunction is obtained by solving the Schrödinger equation. Once the total electron density is known, the electron density under perturbation $\Delta n(r)$ can be obtained by using Eq. (5):

$$n(r) = n_0(r) + \Delta n(r) \qquad (5),$$

where $n_0(r)$ is the electron density at equilibrium state.

Research has shown that when using the charge polarization model to describe the case of an isolated ion in electron free gas, the equation for $F_w$ is identical to that in the semi-ballistic model [15]. The merit of the charge polarization model is that it rests on the charge density contribution calculation and can avoid disentangling the momentum transfer between electrons and ions. Despite the fact that the charge polarization model calculating the electron wind force is shown to be more generalized to realistic systems, the way Bosvieux and Friedel treated the direct force calculation is not relevant, i.e. they considered the direct force to be zero since the screening effect exists. This is not true for alloy systems such as H impurities in metal hosts, i.e. the direct force value is equal to the valence electron of H multiplied by the electric field [15]. Controversies exist for how to properly describe the direct force value to this date [37].

*Pseudopotential-based model*

The way Bosvieux and Friedel treated the individual ion potential $v_o(r)$ was overrated since the repulsive part necessary to cancel the attractive part of $v_o(r)$ near the ion core was not considered. In order to overcome this problem, Sorbello performed a pseudopotential-based analysis to calculate the $F_w$ of the EM driving force [38]. In the pseudopotential method, the potential induced by the core electron and the nuclei is considered as a unified pseudopotential to simplify the complicated description. The pseudopotential of an individual bare ion is $w_0$, and the total pseudopotential of the crystal system is the sum of the individual pseudopotential and the total screened potential (i.e. $w_{total} = \Sigma w_0 + V_{sc}$). Through Eq. (3) to (5), the wind force can then be resolved.

Despite the fact that the pseudopotential-based analysis proposed by Sorbello can be generalized more easily to a real system, the model suffers from arbitrary form factor selection. Metals with a more complicated band structure effect (e.g. *d*- or *f*- electron orbital contributions) will show a huge error range in the effective charge calculation. For example, the error in the calculated effective charge for Cu ranges from 5.61% to 103%; for Ag, it ranges from 14.04% to 64.91%, and for Au, it ranges from 41.92% to 71.62%. On the other hand, the direct force contribution was assumed to be equal to the nominal valence of metals multiplied by the electric field. This assumption may not be universal and again creates controversy. Nevertheless, the analysis done by Sorbello is viewed as a pioneer work to generalize the calculation of the EM driving force.

*Korringa-Kohn-Rostoker* (KKR) *Green's function*

The *Korringa-Kohn-Rostoker* (KKR) method is one of the first principles calculation methods that is wave-function-based. Dekker *et al.* performed the KKR *Green's function* to calculate the electron wind force in pure Al, Cu, Ag, Au, impurities in Al, impurities in Ag, and some 4*d* transition metals in [34]. In [35], Dekker *et al.* systematically calculated the $z^*_{wd}$ in face-centered-cubic and body-centered-cubic metals. In [36], Dekker *et al.* focused on calculating the Al-alloy systems. The electron wind force in these studies was treated by using the polarization model shown in Eq. (3). The alloy system is described with respect to an intermediate system containing a void as a reference system proposed by Lodder [39] in order to depict the saddle point of the diffusion process in EM. The results obtained by Dekker *et al.* appear to be more precise than those obtained using the previous theoretical approach proposed by Sorbello, especially when dealing with noble metals (e.g. Cu, Ag and Au) with complicated band structure. However, a comparison with the experimental data suggests that a quantitative approach to the effective charge fails in terms of some dilute alloy systems, e.g. impurities in Ag [34].

*Recent-proposed models (Machine learning model)*

Recently, Liu *et al.* performed the machine learning method with an experimentally-determined effective charge (i.e. the parameter that set the scales of the EM driving force) data as the training data set, in an attempt to build a model to decipher the effective charge [40]. They found that the effective charge is the function of not only the electrical conductivity of the system, but the electronegativity, periodic table column, and number of $p$ valence electrons. Their results were consistent with the classical understanding but provided far more information related to the effective charge since the classical methods only suggested the effective charge to be a function of valence electrons and electrical conductivity (i.e. $z^* = z_d + \frac{K}{\rho(T)}$). They attempted to predict technologically-relevant host-impurity pairs, such as impurities across the periodic table in the host elements of Al, Ag, Au, Cu, Co and Sn, as shown in Figure 1. Despite the fact that the models seemed to work well, the limitation to the model was the temperature limit (e.g. at only a homologous temperature of 0.9), the composition limit (e.g. only binary systems are valid), and the fact that it can only be applied to metal elements (e.g. non-metal systems such as H may have different underlying physics). The data set may need further refinement to improve the accuracy and generalization for further application.

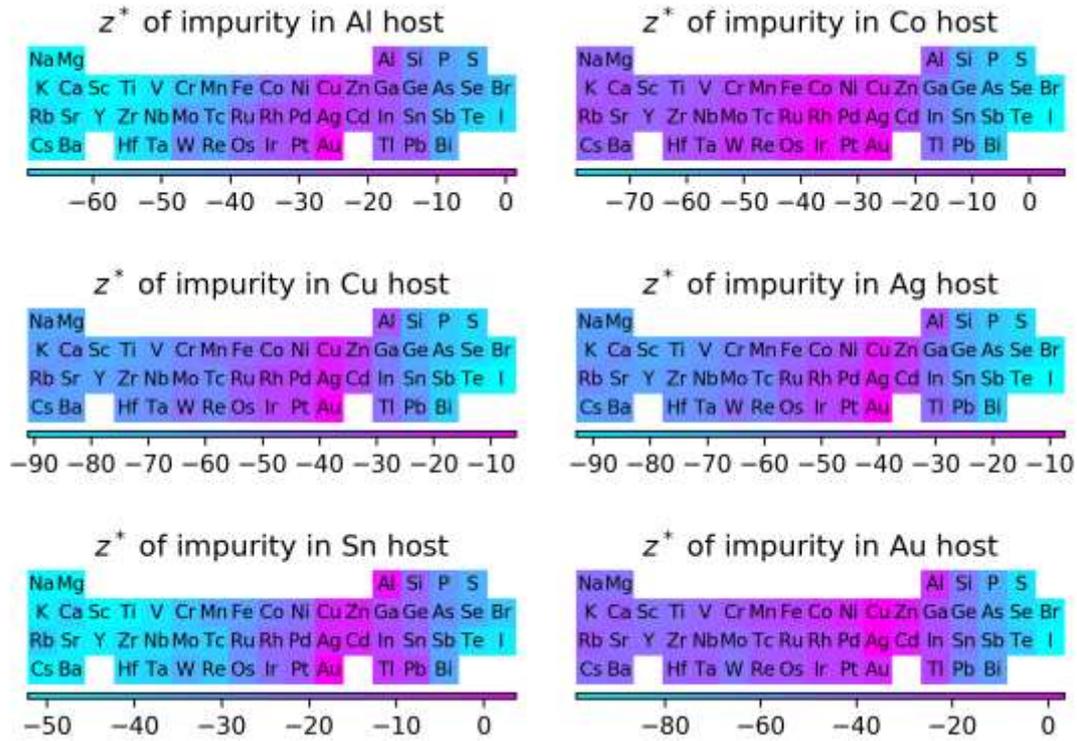

Figure 1. z* of impurities in dilute alloys of Al, Ag, Au, Cu, Co and Sn predicted by the machine learning model. Reprinted with permission from reference [40].

These theoretical models can be well applied to explain EM-induced phenomena such as the formation of voids/hillocks and the polarity effect since they are all directional effects. However, in the following sections, it is shown that only applying the driving force theory for EM cannot explain *non-directional* phenomena, such as the Blech critical product, the phase equilibria change under an electric current, and the non-polarity effect. This suggests that in addition to the diffusion standpoint, other factors governing EM-induced phenomena exist.

**EM-induced stress**

*Blech critical product for EM occurrence and the back stress mechanism*

Blech found that the edge drift of Al strips could only be observed beyond a critical length of the strip under a given current density [41]. The current density multiplied by the critical length was a constant. This product was called the Blech critical length-current-density product or the Blech critical product, where a higher critical product led to more criteria for the occurrence of EM. Blech further found that the Blech critical product of the Al strip increased if the strip was fully covered by a SiN layer, as compared to one without a covered layer. The Blech critical product cannot be well explained by either the semi-ballistic model or the charge polarization model solely since these models does not include any quantities related to the strip length $\ell$. Blech suggested that beside the fact that an electric current will induce force to the ions and cause EM, atomic diffusion itself induced by an electric current will also induce mechanical stress due to atom accumulation before relaxation, which refers to EM-induced stress. Therefore, a so-called back stress existed with the aid of the covered layer. When covered, atoms will accumulate at one side (e.g. anode side). Due to the accumulation of atoms at one side, the vacancy concentration increases at the other side (e.g. cathode) so that a vacancy concentration gradient is established. The side with atom accumulation will develop a compressive stress while the other side with vacancy accumulation will develop a tensile stress. Therefore, a back stress gradient forms that will compensate the EM driving force. The steady-state net flux of diffusion $J$ under electric current is written as Eq. (6) [42]:

$$J = \frac{ND}{kT}[z^*e\rho j - \frac{\partial}{\partial x}(\mu_a - \mu_v)] \qquad (6),$$

where $D$ is the diffusivity, $k$ is the Boltzmann constant, $T$ is the absolute temperature, $z^*$ is the effective charge, $\rho$ is the resistivity, $j$ is the current density, $\mu_a$ and $\mu_v$ are the chemical potential of atoms and vacancies, respectively, and $x$ is the distance along the film in the direction of the electric current. EM-induced atom/vacancy diffusion has been typically assumed to be deposited at grain boundaries. These atoms/vacancies may combine with the grain boundary dislocations and induce dislocation climbing that changes the available lattice sites in the grains (i.e. dislocation climbing is considered to be the source/sinks of lattice sites). It has been assumed that the time span for the local equilibrium between the stress state and the vacancy is far shorter than the long distance diffusion that occurs via grain boundaries. Therefore, Eq. (7) could be established:

$$\mu_a - \mu_v = \mu_0 + \Omega\sigma_{nn} \qquad (7),$$

where $\mu_0$ is a constant, $\Omega$ is the atomic volume, and $\sigma_{nn}$ is the stress normal to the grain boundary. Substituting Eq. (6) with Eq. (7), the steady-state flux $J$ can be written as Eq. (8):

$$J = \frac{ND}{kT}(z^*e\rho j - \frac{d\sigma}{dx}\Omega) \qquad (8),$$

when the electric field-induced driving force was totally compensated by the back stress gradient-induced driving force at steady-state, there was no net flux (i.e. EM ceased), as shown in Eq. (9)-(10):

$$z^*e\rho j = \frac{d\sigma}{dx}\Omega \qquad (9),$$

$$j\ell_{cr} = \frac{\sigma\Omega}{z^*e\rho} \qquad (10),$$

Eq. (10) is well-known as an equation by which to estimate the Blech critical product of EM occurrence. Based on Blech critical product Schreiber further derived a so-called Blech-Schreiber formula to predict the steady-state cathode edge displacement velocity $V$, as shown in Eq. (11) [43]:

$$V = V_d(1 - \frac{L_{th}}{L}) = V_d(1 - \frac{\sigma_{cr}\Omega}{z^*\rho j L}) \qquad (11),$$

where $V_d$ is the displacement velocity for the strip length toward infinity, $L_{th}$ is the threshold strip length for EM occurrence, $L$ is any given strip length, and $\sigma_{cr}$ is the threshold stress for EM occurrence. Despite the fact that debates existed for Eq. (8) to Eq. (11), which will be discussed in the following sections, the Blech critical product is currently still very useful for interconnection design in electronic products. Table I summarizes the critical product of different systems.

Table I. Blech critical product measurement.

| System | Testing structure | Critical product (A/cm) | Temperature (°C) | Reference |
|---|---|---|---|---|
| Sn | Blech structure | 1500 | 63 | [44] |
| Sn-0.7 wt.%Cu | Blech structure | 500 | 63 | [44] |
| Sn-3.0 wt.%Cu | Blech structure | 1580 | 63 | [44] |
| Sn-1.8 wt.%Ag | Solder joint | 30 | 145 | [45] |
| Cu | Blech structure | 900 to 1600 | 175 to 275 | [46] |
| Cu | Dual damascene (processed with a SiCN cap) | 3850±350<br>3760±220<br>4010±200 | 350<br>300<br>250 | [47] |
| Cu | Dual damascene | 3000 | 250 | [48] |
| Cu | Blech structure | 1200 to 2400 | N/A | [49] |
| Cu | Dual damascene | 9000 | 325 | [50] |
| Cu | Single-damascene (passivated with nitride/oxide) | 3940<br>3470<br>2660 | 295<br>350<br>400 | [51] |
| Cu/CNT | Blech structure | 4800 to 6000 | N/A | [49] |
| Al | Blech structure | 1260 | 350 | [41] |
| Al | Blech structure | 901 | 200 | [52] |
| Al-0.5wt.%Cu | Two level structure | 5900 | 210-250 | [53] |

*EM-induced hillocks formation and stress relaxation*

Blech suggested that when the compressive stress at the anode side induced by EM reaches a threshold stress, hillocks will form as the result of stress relaxation. Nevertheless, Eq. (11) assumes the back stress gradient to be linear along the strip length at steady-state, the threshold stress $\sigma_{cr}$ to be constant, and the maximum compressive stress to be built up at the very end of the anode. These assumptions made the hillocks form only at a sharp point (e.g. at the anode end), where the threshold stress was reached. Glickman *et al.* criticized that this was not consistent with the experimental observation since hillock formation is generally found to occur over a finite strip length instead of at a sharp point [54]. Klinger *et al.* [55] and Glickman *et al.* [54, 56] proposed a model to account for hillock formation based on the stress relaxation mechanism. The key assumptions behind their models consisted of the followings: (1) The threshold stress $\sigma_{cr}$ was dependent on the strip length and current density (i.e. the strip length of the hillock formation zone depended on the total strip length and current density), and (2) the relaxation occurred not only via fast dislocation gliding but required time-dependent dislocation movement (e.g. climbing, diffusional creep, *etc.*). Their result experimentally proved that the cathode edge displacement velocity did not follow Eq. (11) (i.e. a negative deviation existed) at higher current densities but approximately followed the model they built based on the diffusional creep mechanism. Even though the authors argued that the threshold stress was intrinsically related to the creep threshold stress, the correct physics underlying the threshold stress induced by EM is still not clear.

*Theoretical frameworks for EM-induced stress evolution*

Even though the aforementioned discussion suggests issues in Eq. (11), the back stress model proposed by Blech is still generally accepted because it is easily comprehended. More accurate theoretical frameworks based on the back stress model to simulate EM-induced stress evolution from the transient state to steady state were further pursued in several works. Kirchheim proposed that when an EM-induced vacancy diffuses to the grain boundary, it causes a volume change in the grain due to the relaxation of its neighboring atoms by $f\Omega$, where $f$ is a relaxation factor, and $\Omega$ is the atomic volume [57]. Due to the strain induced by the vacancy, a new driving force induced by the strain gradient for vacancy diffusion at the grain boundary was established as $J_\sigma = -\frac{DC}{kT}f\Omega\frac{\partial \sigma}{\partial x}$, where $C$ is the vacancy concentration at the grain boundary, and $\frac{\partial \sigma}{\partial x}$ is the stress gradient. The equilibrium vacancy concentration is formulated as $C_e = C_0 exp(\frac{(1-f)\Omega\sigma}{kT})$, where $C_0$ is the equilibrium vacancy concentration in the absent of stress. The total flux of the vacancy is shown in Eq. (12):

$$J = -D\frac{\partial C}{\partial x} + \frac{DC}{kT}z^*e\rho j - \frac{DC}{kT}f\Omega\frac{\partial \sigma}{\partial x} \qquad (12),$$

The first term on right hand side of the equation is due to the driving force of concentration gradient; the second term is due to the electric field, and the third term is due to the vacancy-induced stress gradient. By introducing the annihilation/production of vacancy term (i.e. $\frac{C-C_e}{\tau_s}$, where $\tau_s$ is the relaxation time), the numerical solution for Eq. (12) can be obtained to demonstrate the stress evolution induced by EM.

Korhonen *et al.* proposed an analytical stress evolution model under EM subjected to a confined metal line [58]. They assumed atoms were transported along the grain boundary and are predominately deposited at the grain boundaries under electric current. Following the same assumption made by Blech [42] and Eq. (7), the stress increment caused by the climbing of the grain boundary dislocations in a confined metal line is shown as $d\sigma = -B\frac{dC}{C}$, where $B$ is the applicable modulus, and $\frac{dC}{C}$ is the relative change in the available lattice sites due to dislocation climbing. The stress evolution during EM is thus formulated as Eq. (13):

$$\frac{\partial \sigma}{\partial t} = \frac{\partial}{\partial x}[\frac{D_a B\Omega}{kT}(\frac{\partial \sigma}{\partial x} + \frac{ez^*\rho j}{\Omega})] \quad (13),$$

where $D_a$ is the effective diffusion coefficient of atoms along the grain boundary (i.e. $D_a = \frac{\delta D_{GB}}{d}$, $\delta$ is the grain boundary width, $D_{GB}$ is the diffusion coefficient diffusion coefficient, and $d$ is the grain size. Clement *et al.* further considered the vacancy diffusion mechanism and introduced the vacancy annihilation/creation sink/source term to revisit Korhonen's model, as shown in Eq. (14) [59]:

$$\frac{\partial \sigma}{\partial t} = \frac{D_v C_v B\Omega}{CkT}[\frac{\partial^2 \sigma}{\partial x^2} + (\frac{\Omega}{kT}\frac{\partial \sigma}{\partial x} + \frac{z^*e\rho j}{kT})\frac{\partial \sigma}{\partial x}] \quad (14),$$

where $D_v$ is the vacancy diffusion coefficient, $C_v$ is the equilibrium vacancy concentration, and $C$ is the lattice site concentrations. Clement *et al.* suggested that Korhonen's model is only valid when the EM-induced maximum stress is relatively small. Park *et al.* followed Korhonen's model but further discussed the effect of the stress on the atomic diffusivity [60]. The stress-dependent diffusivity is shown as $D' =$

D $\exp[(\frac{\Omega}{kT} + \frac{1}{B})\sigma]$. Lloyd summarized the analytical models for evaluating the EM-induced stress including the thermal stress effect [61]. Sarychev *et al.* further provided an analytical model to solve three dimensional EM-induced stress in analog to the thermal stress [62].

*Debates for the models simulating the EM-induced stress evolution*

The vital question for models simulating the EM-induced stress evolution is related to determining how the stress will build up without the external confinement (e.g. oxide layer, passivated layer, *etc.*) necessary to help the atoms accumulate. Tu argued that the back stress model might not be valid for cases of non-covered strips (e.g. Au) since atom accumulation will be instantly relaxed, and no stress will be induced [63]. It also might not be valid for cases undergoing surface diffusion (e.g. Cu) since the atoms accumulating on the surface cannot build stress for the bulk. The boundary condition of the covered layers seems to make the back stress model and the associated theoretical frameworks phenomenological and not general to all cases. To be more specific, this gives rise to five points that are unclear: (1) How will the stress build up without an external constraint? (2) How will the stress relax (e.g. via dislocation glide, climb, creep, *etc.*)? (3) How fast is the stress build-up compared with the stress relaxation? (4) What is the physics underlying the critical stress that causes electric current-induced deformation? (5) Is the critical stress a constant, or does it depend on the current density or the strip length?

On the other hand, Eq. (10) suggests that the product is only valid for a short strip length. Typically, the upper bound (i.e. the threshold) of the EM-induced stress is considered to be the yield strength ($\sigma_y$) of the material [63]. The yield stress of metals is generally in the hundred MPa scale. For instance in Al, $\sigma_y = 110$ MPa, $\Omega = 1.67 \times 10^{-29}$ $m^3, z^* = -13, \rho = 2.6 \times 10^{-8}$ $m\Omega$, and the typical lower bound of current density for EM occurrence in Al is $10^4$ $A/cm^2$. Following Eq. (10), the upper bound for the threshold strip length yields 340 $\mu m$, which agrees with the typical strip length scale used in the experiment [41, 52, 53]. It is likely that above this sub-micron length scale Eq. (10) would no longer be valid. This seems true in the experimental observation made by Lin *et al*, where EM was found to occur in a 2-cm-long Cu strip subjected to a current density of *ca.* $7.5 \times 10^5$ A/cm², which corresponds to a critical product of $1.5 \times 10^6$ A/cm. This seems way larger than the conventional observation, e.g. 900 [46] to 9000 [50] A/cm. This may suggest a missing part exists in the Blech critical product theoretical framework.

*EM-induced lattice deformation measurement*

A series of *in situ* synchrotron radiation (SR)-based current stressing experiments were conducted to gain more insights into the EM-induced stress evolution from experimentally measuring the lattice deformation under an electric current. The deformation include lattice strain evolution [20-24], dislocation formation [25, 26, 64-66], grain rotation [64, 66, 67], and sub-grain formation (i.e. polygonization) [25, 26, 64]. A quantitative study showed the resolved shear stress (based on the FCC metals slip system) at the location where atoms accumulate (e.g. hillocks were found) to be 10 times larger than at a location without atom accumulation [65]. The location where atoms accumulate was suggested to have higher dislocation density [65]. Valek *et al.* studied the early stage (i.e. before the void/hillock formation induced by EM could be observed) of crystal plasticity change induced by the electric current of Al, as shown in Figure 2 [25]. They found very similar results to those in [64], where individual grains underwent bending induced by the electric current and thereby introduced a preferred-oriented dislocation formation (a dislocation density of $3 \times 10^9$ /cm$^2$) and formation of small angle grain boundaries in the direction of the electron flow (presumably via dislocation climbing) before the void/hillock formation. Similar results were suggested by Budiman *et al.*, who studied Cu at the early stage of an electric current [26], and the dislocation density was found to be $3 \times 10^9$ /cm$^2$.

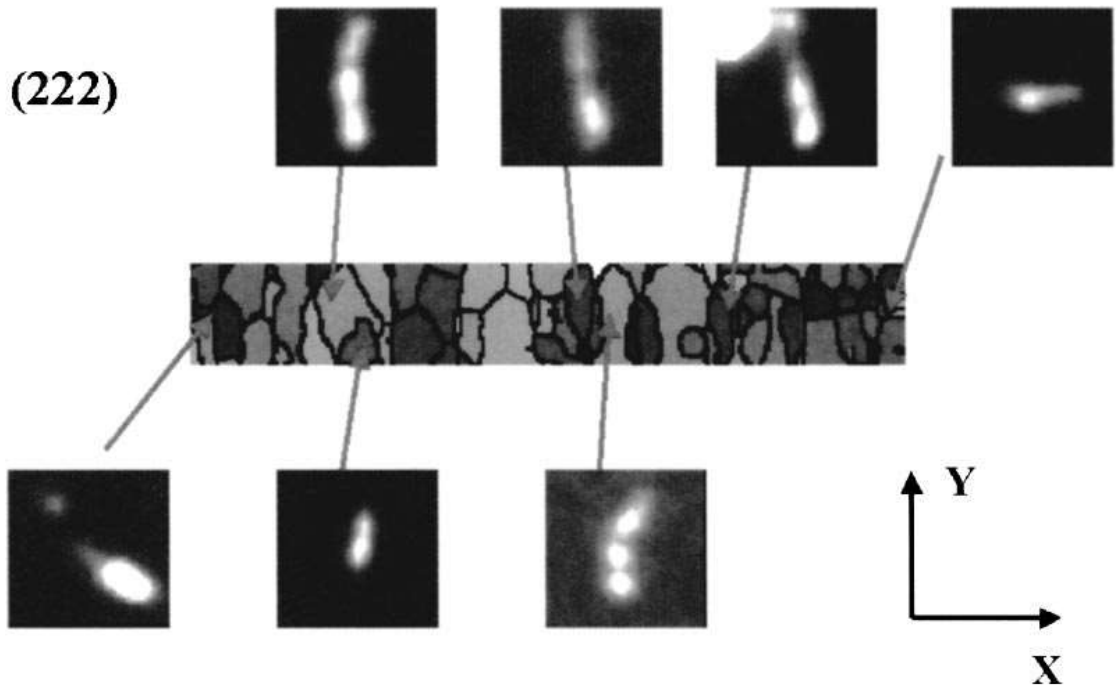

Figure 2. The (222) plane peak profile at different grains along the Al strip subjected to electric current. Peak streaking, rotation, and broadening were found under electric current, indicating the plastic deformation at the early-stage of current-stressing. Reprinted with permission from reference [25].

Here, an effort is made to highlight the experimental works done at the early stage of EM [24-26], which experimentally suggested that before the void/hillock formation, crystal plasticity including the dislocation formation was found. A plausible explanation for these findings was that the plasticity resulted from the stress developed by the atom/vacancy accumulation, which was discussed in the aforementioned section. However, how the atom/vacancy accumulation will induce stress without covered layers is not clear. A more rigid quantitative model and mechanistic study to correlate the electric current and early-stage stress state is required. Nevertheless, these works provided a different insight suggesting that the way to view EM-induce failures (e.g. void/hillock formation) is not just a kinetics problem but also involves the mechanical stress.

An additional insight is that Lin *et al.* found that voids/hillocks formed at a homologous temperature of 0.28 under current stressing for 1680 s [24]. Under this condition, very low kinetics was suggested (i.e. where a surface diffusion coefficient of $10^{-16}$ m$^2$/s, an absolute temperature of 373 K, an effective charge of -5.5, resistivity of $1.68 \times 10^{-8}$ m·Ω, a current density of $10^{10}$ A/m$^2$ yielded a self-diffusion velocity of *ca.* 0.001 nm/s). From this velocity, it was not possible to find a hillock size of *ca.* 1 μm after 1680 s of current-stressing. The classical models for the EM driving force seem to be missing some aspects of the low kinetics. With an understanding of the electric current-induced plastic deformation found at the early-stage, one plausible explanation for this issue is that an electric current will induce dislocation formation. Either the stress relaxation may result in hillock formation, as shown schematically in Figure 3 [24], or the dislocations may provide a fast diffusion path [25, 26, 64-66]. It is also likely that a high resistivity location (e.g. grain boundary) may provide local Joule heating and extra electron wind force that dominates the diffusion process. Overall, only applying the kinetics perspective seems inadequate to elucidate the observed phenomenon.

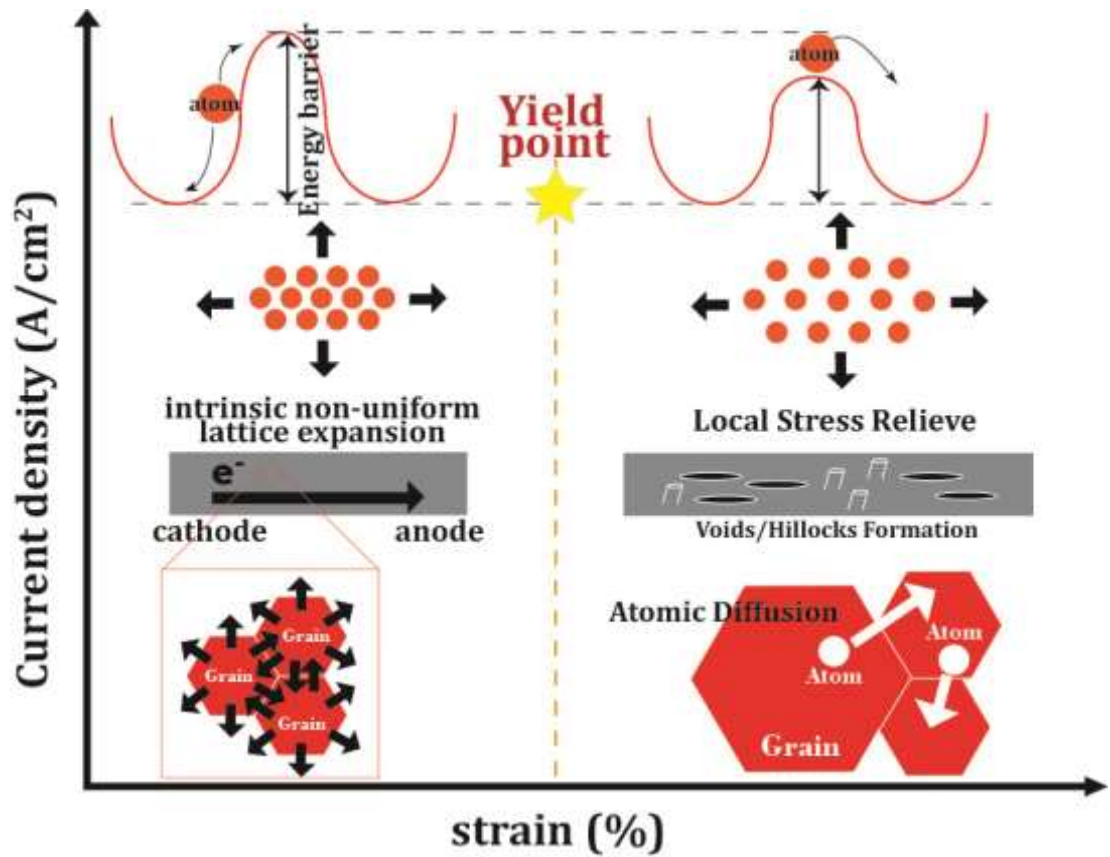

Figure 3. Schematic diagram of the stress relaxation induced voids/hillocks formation subjected to electric current. Reprinted with permission from reference [24].

**Correlation between the mechanical properties of materials and EM occurrence**

The aforementioned discussions point out that EM will induce mechanical stress due to atom accumulation/depletion before relaxation occurs. This infers that EM is likely to be the result of the electric current-induced crystal plasticity. When the stress reaches a "yield" strength of materials, plastic deformation, including the void/hillock formation, and crystal plasticity will be found. If this is true, it is then sensible to ask one question regarding the correlation between EM and mechanical stress － does EM occurrence correlate to a material's mechanical properties? To be more specific, do materials with stronger mechanical properties tend to have stronger EM-resistance? The answer to these questions seem positive when observing what people have done in regard to enhancing the EM-resistance of materials. For example, (111)-textured face-centered cubic (fcc) materials have stronger EM-resistance than non-textured ones [10, 68-70], while (111)-textured fcc materials have been found to have a maximum Young's modulus compared to other planes. Precipitation of $Al_2Cu$ existing in an Al-Cu alloy was found to have a longer mean time to failure (MTF) [71, 72], and precipitation has been found to harden materials. EM occurrence was found to be delayed when fully-covered layers were present, e.g. anodization [73], and passivated layers [50]. These "rigid walls" may suppress the lattice deformation induced by an electric current. Therefore, materials are more prone to being retained in the elastic deformation region [24]. Though one may argue that precipitation or segregation found at the grain boundary retards EM-induced diffusion [74], the reason for enhancement of EM-resistivity by textured film, covered film, and the presence of precipitation is not easily understood via pure diffusion theory. These indirect correlations likely suggest that EM occurrence is strongly related to the mechanical properties of materials.

**Electroplastic (EP) effect**

The aforementioned discussion suggests that the mechanical stress/mechanical properties of materials may be one of the key factors that can be used to evaluate the EM effect, especially when speaking of EM-induced failure such as formation of voids/hillocks, the Blech effect, EM-induced stress, *etc*. However, most EM studies have solely focused on the atomic diffusion aspect (i.e. elucidating the EM-related phenomena via various diffusion theories). Very few EM studies have provided direct experimental evidence to prove that EM occurrence is mechanistically related to the mechanical properties of materials [63, 75]. The very direct correlation between an electric current and plastic deformation has been found for a related effect called the electroplastic (EP) effect, which refers to the plastic deformation induced by current stressing [76-81]. The theoretical EP model is generally accepted as an electron-dislocation interaction, but this is still being debated up to the present time [79]. The most well-known phenomenon induced by EP is where the electric current decreases the flow stress but no change is found in the elastic stress, as shown in Figure 4 [82-84]. The recrystallization temperature was found to decrease under current stressing (i.e. to accelerate the occurrence of recrystallization) [85]. The grain size of materials has been found to be refined under current stressing [86-88]. Even though scientists usually consider the EP effect to be different from that of EM by suggesting the model for the EM driving force is merely one way to explain EP (i.e. the local Joule heating is considered as one of the possible mechanisms for EP) [89], it is still very difficult to separate EM from EP when current-stressing materials. Up to the present time, no studies have directly proven that these two effects are parallel. Based on our understanding and the aforementioned discussions, it is likely that EM is part of the EP effect framework, or EM shares the intrinsic electric current-induced plastic deformation induced by the EP effect. In this context, the crystal plasticity found in the early-stage of the current stress is likely to result from the EP effect. EP effect might be the reason why electric current will induce stress without external covering layer.

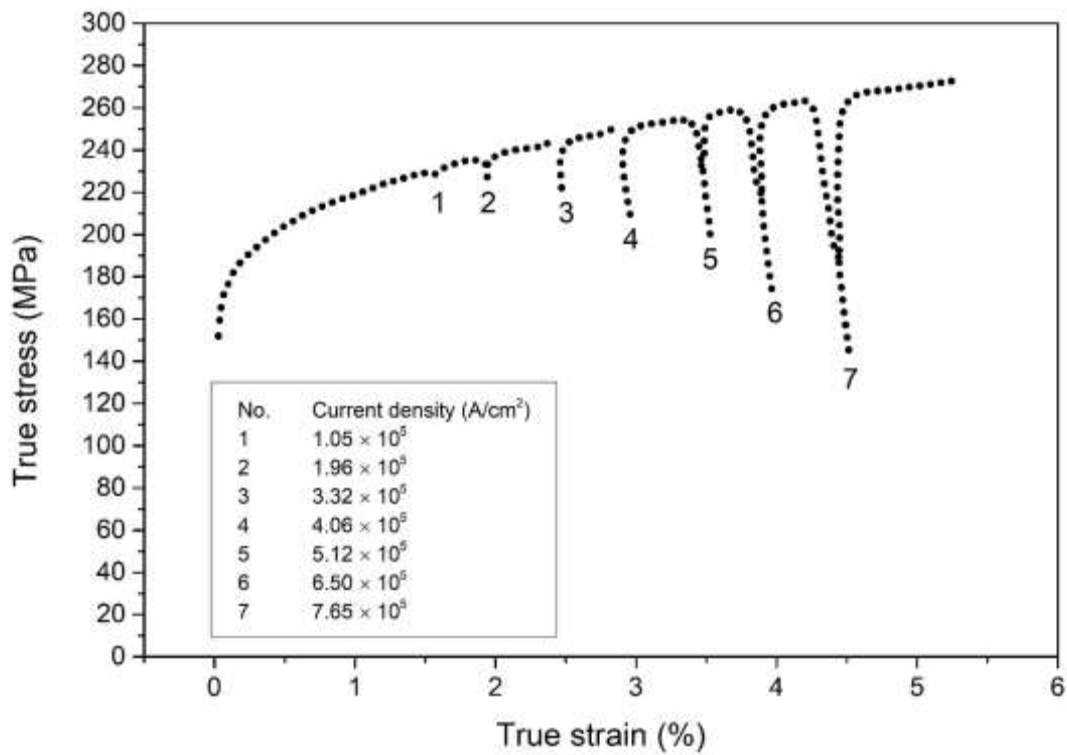

Figure 4. The true stress vs. true strain relation of Ti subjected to pulse direct current. At each pulse, the true stress dropped. Data was taken from [84].

**Electric current-induced phase equilibria change**

Electric current has been found to change the phase equilibria of materials [17, 27, 89-102]. Conrad provided a thorough discussion of phase transformation under an electric current, including the interfacial reaction, precipitation, crystallization, and recrystallization of metals [89]. Guan *et al.* reviewed and discussed the electric current-induced phase transformation effect [81]. Dolinsky and Elperin discussed the thermodynamics of phase transition [103] and nucleation [104] under electric current. They showed that additional work is required to form a nucleation with lower electrical conductivity. In other words, an additional energy term is added in the Gibbs free energy change of a given phase transformation due to the change in the current distribution when a nucleus forms. Jiang *et al.* performed a thermodynamics analysis of the dissolution of the beta phase in the Mg-Al-Zn alloy under electric current following Dolinsky and Elperin's work, and the analytical results agreed with their experimental results [18]. Lin *et al.* used the *ab initio*-aided CALPHAD method to explain the supersaturation of Pb-Sn induced by an electric current [105]. The relationship between compressive stress and current density was empirically established by correlating that the Sn whisker growth rate was the same under the associated electric current or compressive stress [106]. The modeling results showed that when the current density is higher than a critical value of *ca.* $2.5 \times 10^4$ A/cm$^2$, the phase boundaries of the Pb-Sn system are changed by electric current stressing, as shown in Figure 5. The simulation results agreed well with the experimental observations [17]. However, the correlation between the stress and current density in the study was only empirical and lacked a physical interpretation.

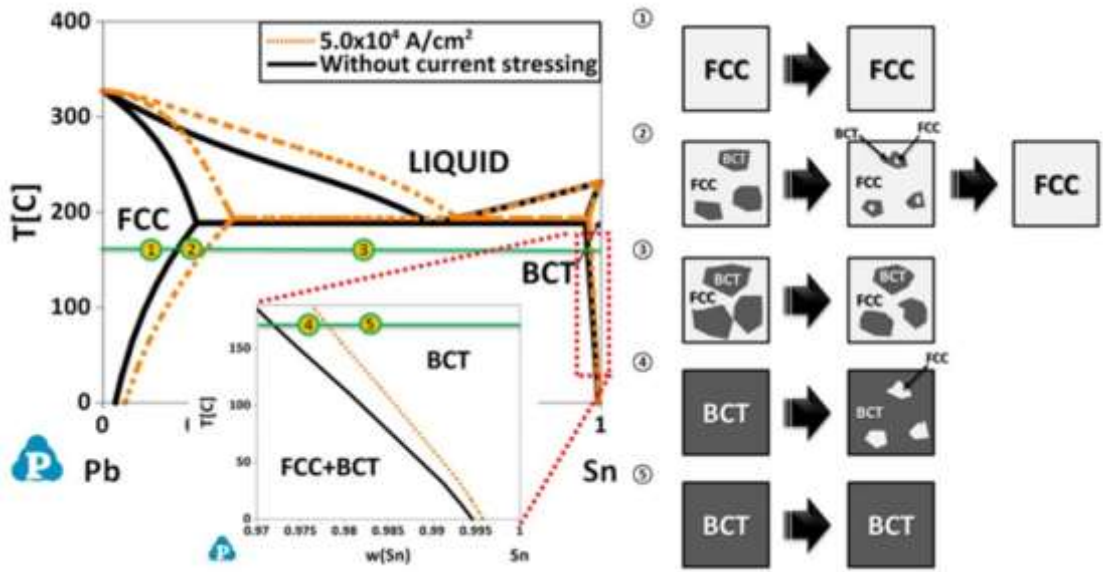

Figure 5. The phase boundary and the phase relation of Pb-Sn binary system change under electric current. Reprinted with permission from reference [105].

The phase equilibria change is a *non-directional* effect and thus cannot be well explained by the classical theoretical models of EM. However, if an electric current intrinsically induces lattice plasticity, and it does not relax to form voids/hillocks or any other relevant failures, the stored strain energy will be likely to contribute to the Gibbs free energy change and the driving force in the phase transformation. In general the molar Gibbs free energy of any given phase changed by electric current-induced strain can be described using Eq. (15):

$$G_m' = {}^0G_{ref} + {}^{id}G_m + {}^{ex}G_m + {}^{strain-ex}G_m \qquad (15),$$

where ${}^0G_{ref}, {}^{id}G_m, {}^{ex}G_m, {}^{strain-ex}G_m$ are the are the reference energy of the constituent components, and the molar Gibbs free energy, ideal mixing Gibbs free energy, excess Gibbs free energy due to non-ideal mixing, and excess Gibbs free energy due to the electric current-induced strain, respectively. Therefore, the addition of the ${}^{strain-ex}G_m$ term might result in an abnormal supersaturation phenomenon, and would also be the origin of the stress introduced to the CALPHAD model proposed by Lin *et al.* [105].

**Polarized and non-polarized effect**

Interfacial reactions are found to be changed by the EM effect [107, 108], including the polarity and non-polarity effects. In a given reaction couple (e.g. an A/B/A sandwich structure) forming interface layers (e.g. A/L/B/L/A, where *L* denotes the interface layer), the flux of dominant diffusion species *a* ($J_a$) under a direct electric current can be denoted as Eq. (16):

$$J_a = -\frac{D_a N_a}{RT}\left(RT\frac{\partial \ln N_a}{\partial x} \pm z_a^* eE\right) \qquad (16),$$

where $N_a$ is the mole fraction of species *a,* and *x* is the thickness of the interface layer formed at the interface. If the dominant diffusion species *a* diffuses in the same direction as the transporting electron flow to form the interface layer, the second term in the parenthesis on the right hand side of Eq. (16) will be $+z_a^* eE$. If the diffusion direction is opposite to that of the electron flow, the second term will be $-z_a^* eE$. Eq. (16) clearly indicates that in the case of the two interface layers in the given A/B/A sandwich structure under an electric current, one will be enhanced and the other one will be suppressed comparing to the reaction couple without current stressing. This phenomenon is called the polarity effect. EM theories of the semi-ballistic model or the charge polarization model can explain the polarity effect effectively. Table II summarizes the experimental results for the polarity effect found in several reaction couple systems.

However, the direct current has also been found to have either no effect on the growth rate of the interface layers, or to cause a symmetrical enhancement/suppression of the interface layers at both the cathode and anode side. These results were in contrast to the usual polarity effect on the asymmetrical growth rate of the interface layers, and were independent of the electric current direction. Therefore, the phenomenon showing symmetrical enhancement/suppression of the interface layer under an electric current is called the non-polarity effect. Eq. (16) then fails to explain the non-polarity effect since it is a *non-directional* effect. A series of models was proposed to explain the non-polarity effect, but are plausible [109, 110] and phenomenological [111]. A better understanding of how investigate the underlying physics of the non-polarity effect is still being pursued. Table III summarizes the experimental results showing no effects of a direct current on the interfacial reaction and the results for the non-polarity effect. It is worth mentioning that AC or a reversing current have also been found to induce the non-polarity effect, as summarized in Table III.

The non-polarity effect is likely to result from the phase equilibria changed by electric current. To be more specific, the chemical potential of a given species is changed by electric current as suggested by Eq. (15). For example, the diffusion-controlled $NiBi_3$ interface layers at the cathode and the anode side of Bi/Ni reaction couple were found to be both enhanced under electric current [19]. If the chemical potential of the $NiBi_3$ phase is changed by the electric current due to the presence of strain energy, then the chemical potential gradient of the dominant diffusion species controlling the growth of $NiBi_3$, which is Bi in this case, will be changed simultaneously. If the chemical potential gradient of Bi is enhanced by the electric current due to the strain energy, the diffusion velocity of Bi will be enhanced, and the diffusion will be independent of the direction of the electron flow [112]. Therefore, $NiBi_3$ growth will be accelerated. This is likely to be true for $NiBi_3$ growth since it is a diffusion-controlled reaction process (i.e. the reaction between Bi and Ni is fast enough so that the $NiBi_3$ phase forms immediately when Bi reaches Ni). In short, it is likely that the non-polarity effect originates from the change in the electric current-induced phase stability, which changes the driving force of diffusion in the meantime.

Table II. Summary of the experimental results for the polarity effect found in several reaction couple systems. Adapted from Ref. [111].

| System (Temperature) | Current density (A/cm$^2$) | *IMC Growth | | Reference |
|---|---|---|---|---|
| | | Cathode | Anode | |
| Sn/Ni (160-200°C) | 500 | E | S | [13] |
| Sn/Ag (120-200°C) | 500 | E | S | [113] |
| Sn-3.5wt%Ag/Ni (160-200°C) | 500 | E | S | [114] |
| Sn-0.7wt%Cu/Ni (160-200°C) | 500 | E | S | [114] |
| Cu/Sn-3.8wt%Ag-0.7wt%Cu (180°C) | 10$^3$ to 10$^4$ | S | E | [12] |

*E: Enhanced; S: Suppressed.

Table III. Summary of the experimental results showing no effects of direct current upon the interfacial reaction and results for the non-polarity effect. Adapted from Ref. [111].

| System (Temperature) | Current density (A/cm$^2$) | *IMC Growth | | Reference |
|---|---|---|---|---|
| | | Cathode | Anode | |
| Sn/Cu (200°C) | 500 | N | N | [13] |
| Zn/Ni (150,200°C) | 300 | N | N | [19] |
| Bi/Ni (185,200°C) | 300 | N | N | [19] |
| Al/Ni (400°C) | 500,1000 | E | E | [109] |
| Bi/Ni (150-170°C) | 300 | E | E | [19] |
| Al/Au (400-500°C) | 1000 | E | E | [110] |

| | | | | | |
|---|---|---|---|---|---|
| Ni/Ti (625~850°C) | 0 to 2546 | E | E | | [115] |
| Ag/Zn (300,350°C) | 0 to 764 | E | E | | [116] |
| Ni-P/Sn-3.5%Ag | 1000 | S | S | | [117] |
| $^\$$Al/Cu (200-525°C) | 500 to 1300 | E | E | | [118] |
| $^\$$Sn/Ag (160°C) | 500 | E | E | | [119] |
| $^\$$Sn/Ni (180-200°C) | 500 | E | E | | [119] |

*N: No effect; E: Enhanced; S: Suppressed.

$^\$$Under alternating current or reversing current.

**Conclusion**

This review article provides a thorough review on the EM-related phenomena, especially for *non-directional* effects, EM-related theoretical models, and the relevant EP effects. A broad picture of electric current-induced phase stability changes is proposed, which could complement the missing aspects of the existing EM theories for the *non-directional* phenomena. Main concerns in the existing EM theories are summarized as follows: (1) Although the general picture of the EM theory, *i.e.*, the charge polarization model, has been established, it is still not fully complete due to the fact that the $z_d$ value cannot be easily simulated; (2) the classical theories are not fully applicable to explain the non-directional changes of phase equilibria induced by electric currents; (3) the existing theoretical frameworks were developed based on experiments with external covers, such as native oxides or passivated layers, so the stress could evolved; however, it is not clear why the stress could be built up without external covers. The origin of the EM-induced stress and early-stage crystal plasticity is not clear.

By introducing the electric current-induced plasticity, or the EP effect to the EM effect, the aforementioned electric current-induced peculiar phenomena could be comprehended. In the case of electric current-induced crystal plasticity, the additional strain energy could contribute to the changes in Gibbs free energy of phase transformation. Therefore, the phase stability and interfacial reactions are expected to be changed by electric currents. The EP effect might also be the reason for early-stage lattice deformation observed in the EM studies. The missing part of EM theory to elucidate the *non-directional* effect is proposed in the review article. Further investigations for establishing a complete theory for the effect of electric currents upon materials are needed.

**Acknowledgement**

The authors gratefully acknowledge the financial support from the Ministry of Science and Technology (MOST) in Taiwan (MOST 103-2221-E-006-043-MY3 and 106-2628-E-006-002-MY3).

**Authors' contributions**

Y.C.L collected important papers and made first draft of the article with the guidance of S.K.L. S.K.L modified and revised for publication. All authors read and approved the final manuscript.

**Table Caption:**

Table I. Blech critical product measurement.

Table II. Summary of the experimental results for the polarity effect found in several reaction couple systems. Adapted from Ref. [111].

Table III. Summary of the experimental results showing no effects of direct current upon the interfacial reaction and results for the non-polarity effect. Adapted from Ref. [111].

**Figure Captions:**

Figure 1. z* of impurities in dilute alloys of Al, Ag, Au, Cu, Co and Sn predicted by the machine learning model. Reprinted with permission from reference [40].

Figure 2. The (222) plane peak profile at different grains along the Al strip subjected to electric current. Peak streaking, rotation, and broadening were found under electric current, indicating the plastic deformation at the early-stage of current-stressing. Reprinted with permission from reference [25].

Figure 3. Schematic diagram of the stress relaxation induced voids/hillocks formation subjected to electric current. Reprinted with permission from reference [24].

Figure 4. The true stress vs. true strain relation of Ti subjected to pulse direct current. At each pulse, the true stress dropped. Data was taken from [84].

Figure 5. The phase boundary and the phase relation of Pb-Sn binary system change under electric current. Reprinted with permission from reference [105].